\documentclass[epj]{webofc}
\usepackage[utf8]{inputenc}
\usepackage[varg]{txfonts}   
\usepackage{bm}
\usepackage{xcolor}
\definecolor{darkred}{rgb}{0.4,0.0,0.0}
\definecolor{darkgreen}{rgb}{0.0,0.4,0.0}
\definecolor{darkblue}{rgb}{0.0,0.0,0.4}
\usepackage[bookmarks,linktocpage,colorlinks,
    linkcolor = darkred,
    urlcolor  = darkblue,
    citecolor = darkgreen]{hyperref}
\usepackage{subfigure}
\wocname{EPJ Web of Conferences}
\woctitle{Lattice2017}
\newcommand{\nn}{\nonumber}

\newcommand{\svec}{\bm{s}}
\newcommand{\sveca}{\bm{s}_a}

\newcommand{\Pvec}{\bm{P}}

\newcommand{\dvec}{\bm{d}}

\newcommand{\ket}[1]{\vert #1\rangle}

\newcommand{\me}[3]{\langle #1\vert\ #2\ \vert #3\rangle}

\newcommand{\beq}{\begin{equation}}
\newcommand{\eeq}{\end{equation}}
\newcommand{\beqy}{\begin{eqnarray}}
\newcommand{\eeqy}{\end{eqnarray}}
\newcommand{\beqqy}{\begin{eqnarray*}}
\newcommand{\eeqqy}{\end{eqnarray*}}

\newcommand{\Ecm}{E_{\rm cm}}

\newcommand{\qcma}{\bm{q}_{{\rm cm},a}}

\begin{document}
%
\selectlanguage{english}
\title{%
Scattering from finite-volume energies including higher partial waves 
and multiple decay channels
}
\author{%
\firstname{Ruair\'i} \lastname{Brett}\inst{1}\and
\firstname{John} \lastname{Bulava}\inst{2}\fnsep\thanks{Speaker, \email{bulava@imada.sdu.dk}} \and
\firstname{Jacob}  \lastname{Fallica}\inst{1}\and
\firstname{Andrew}  \lastname{Hanlon}\inst{3}\and
\firstname{Ben}  \lastname{H\"orz}\inst{4}\and
\firstname{Colin}  \lastname{Morningstar}\inst{1}\and
\firstname{Bijit}  \lastname{Singha}\inst{1}
}
\institute{%
Department~of~Physics, Carnegie~Mellon~University, Pittsburgh, PA~15213, USA
\and
Dept.~of Mathematics~and~Computer~Science and CP3-Origins, 
     University of Southern Denmark, Campusvej 55, 5230 Odense M, Denmark
\and
Dept.~of Physics and Astronomy, University of Pittsburgh, 
     Pittsburgh, PA 15260, USA
\and
PRISMA Cluster of Excellence and Institute for Nuclear Physics, 
     University of Mainz, Johann Joachim Becher-Weg 45, 55099 Mainz, Germany
}
\abstract{%
    A new implementation of estimating the two-to-two $K$-matrix from finite-volume
energies based on the Luescher formalism is described.  The method includes higher 
partial waves and multiple decay channels, and the fitting procedure properly 
includes all covariances and statistical uncertainties.  The method is also simpler 
than previously used procedures.  Formulas and software for handling total spins 
up to $S=2$ and orbital angular momenta up to $L=6$ are presented.  
}
\maketitle
\section{Introduction}\label{intro}

A key goal in lattice QCD is to determine the spectrum of hadrons.
Lattice simulations can only calculate the energies of stationary
states in finite volume.  Since excited hadrons are unstable resonances,
their masses and decay widths must be deduced from the finite-volume 
energies obtained in lattice QCD using rather complicated formulas, 
developed over several decades in Refs.~\cite{Luscher:1990ux,Rummukainen:1995vs,
Kim:2005gf,Briceno:2014oea}, among others.
This talk reports on work completed in Ref.~\cite{Morningstar:2017spu} to 
provide explicit formulas, software, and new fitting implementations 
for carrying out two-particle scattering studies using energies
obtained from lattice QCD.  In our first tests, we 
incorporate the $L=3$ and $L=5$ partial waves in the decay of the $\rho$-meson 
to two pions and find their contributions to be negligible in the elastic 
energy region.

\section{Quantization condition}

Let $\Pvec=(2\pi/L)\dvec$, where $\dvec$ is a vector of integers, denote
a total momentum in an $L^3$ spatial volume with periodic boundary conditions.
The center-of-momentum energy $\Ecm$ is related to the lab frame energy $E$ by
\beq
   \Ecm = \sqrt{E^2-\Pvec^2},\qquad \gamma =\frac{E}{\Ecm}.
\eeq
Let $N_d$ denote the number of two-particle channels that are open,
and denote the masses and spins of the two scattering particles in
channel $a$ by $m_{ja}$ and $s_{ja}$, respectively, for $j=1,2$.
In each channel, we can define a quantity $\qcma^2$ by
\begin{equation}
   \sqrt{\qcma^2+m_{1a}^2}+\sqrt{\qcma^2+m_{2a}^2}=\Ecm.
\end{equation}
Since $\Ecm^2=E^2-\Pvec^2$ must be real, a real solution for $\qcma^2$
exists if $\vert\Ecm^2\vert \geq \vert m_{1a}^2-m_{2a}^2\vert$, then 
we can calculate the following quantities in each channel:
\begin{eqnarray}
   \qcma^{2} &=& \frac{1}{4} \Ecm^{2}
   - \frac{1}{2}(m_{1a}^2+m_{2a}^2) + \frac{(m_{1a}^2-m_{2a}^2)^2}{4\Ecm^{2}},\\
   u_a^2&=& \frac{L^2\qcma^2}{(2\pi)^2},\qquad
 \sveca= \left(1+\frac{(m_{1a}^2-m_{2a}^2)}{\Ecm^{2}}\right)\dvec.
\end{eqnarray}
The total energy $E$ is then related to the dimensionless unitary scattering $S$-matrix
through the quantization condition\cite{Luscher:1990ux,Rummukainen:1995vs,Kim:2005gf,Briceno:2014oea}:
\beq
   \det[1+F^{(\Pvec)}(S-1)]=0.
\label{eq:quant1}
\eeq
In an orthonormal basis of states, each labelled by $\ket{Jm_JLS a}$, where $J$ is
the total angular momentum of the two particles in the center-of-momentum frame, $m_J$ is 
the projection of the total angular momentum onto the $z$-axis, $L$ is the orbital angular 
momentum of the two particles in the center-of-momentum frame (not to be confused with the
lattice length here), and $S$ is the total spin of the two particles (not the scattering matrix). 
The index $a$ is generalized to refer to species, the spins $s_1,s_2$, intrinsic 
parities $\eta^P_1,\eta^P_2$, isospins $I_1, I_2$, isospin projections $I_{z1}, I_{z2}$, 
and possibly $G$-parities $\eta_1^G, \eta_2^G$ of particle 1 and 2. 
The $F^{(\Pvec)}$ matrix in this basis is given by
\beqy
&&\langle  J'm_{J'}L'S'a'\vert F^{(\Pvec)}\vert Jm_JLSa\rangle
=\delta_{a'a}\delta_{S'S}\ \frac{1}{2}
\Bigl\{\delta_{J'J}\delta_{m_{J'}m_J}\delta_{L'L}\nn\\
&&\qquad + \langle J'm_{J'}\vert L'm_{L'} Sm_{S}\rangle
\langle Lm_L Sm_S\vert Jm_J\rangle
W_{L'm_{L'};\ Lm_L}^{(\Pvec a)}\Bigr\},
\eeqy
where $\langle j_1m_1 j_2m_2\vert JM\rangle$ are the familiar Clebsch-Gordan coefficients,
and the $W^{(\Pvec a)}$ matrix elements are given by
\beqy
-iW^{(\Pvec a)}_{L'm_{L'};\ Lm_L} 
&=& \sum_{l=\vert L'-L\vert}^{L'+L}\sum_{m=-l}^l
   \frac{ {\cal Z}_{lm}(\svec_a,\gamma,u_a^2) }{\pi^{3/2}\gamma u_a^{l+1}}
\sqrt{\frac{(2{L'}\!+\!1)(2l\!+\!1)}{(2L+1)}}
\langle {L'} 0,l 0\vert L 0\rangle
\langle {L'} m_{L'},  l m\vert  L m_L\rangle.
\label{eq:Wdef2}
\eeqy
The Rummukainen-Gottlieb-L\"uscher (RGL) shifted zeta functions 
${\cal Z}_{lm}$, introduced in Refs.~\cite{Luscher:1990ux,Rummukainen:1995vs},
are evaluated as detailed in Ref.~\cite{Morningstar:2017spu}.

\section{The K-matrix and box matrix}

Eq.~(\ref{eq:quant1}) is a single relation between a lab-frame finite-volume 
energy $E$ and the entire $S$-matrix. This single relation is insufficient
to extract all of the $S$-matrix elements at energy $E$.  We proceed by the
usual method of approximating the $S$-matrix elements using physically motivated 
functions of the energy $E$ involving a handful of parameters.  Values
of these parameters can then hopefully be estimated by appropriate fits
using a sufficiently large number of different energies. 

The $S$-matrix in Eq.~(\ref{eq:quant1}) is dimensionless and unitary.
Since it is easier to parametrize a real symmetric matrix than a unitary 
matrix, one usually employs the real and symmetric 
$K$-matrix\cite{Wigner:1946zz,Wigner:1947zz}, defined by
\beq
   S = (1+iK)(1-iK)^{-1} = (1-iK)^{-1}(1+iK). 
\eeq
Rotational invariance implies that the $K$-matrix 
must have the form
\beq 
  \langle J'm_{J'}L^\prime S^\prime a'\vert\ K
\ \vert Jm_JLS  a\rangle = \delta_{J'J}\delta_{m_{J'}m_J}
 \ K^{(J)}_{L'S'a';\ LS a}(\Ecm),
\label{eq:Kbasis}
\eeq
where $a',a$ denote other defining quantum numbers, such as channel, and
$K^{(J)}$ is a real, symmetric matrix that is independent of 
$m_J$. Invariance under parity also gives us that
\beq
   K^{(J)}_{L'S'a';\ LSa}(\Ecm)=0\quad\mbox{when 
  $\eta^{P\prime}_{1a'}\eta^P_{1a}\eta^{P\prime}_{2a'}\eta^P_{2a}(-1)^{L'+L}=-1$},
\label{eq:Kparity}
\eeq
where $\eta_{ja}^P$ denotes the intrinsic parity of particle $j$ in
the channel associated with $a$.
The multichannel generalization\cite{Ross:1961aa,deSwart:1962aa,Burke:2011}
of the effective range expansion is
\beq
 K^{-1}_{L'S'a';\ LSa}(\Ecm)=q_{{\rm cm},a'}^{-L'-\frac{1}{2}}
 \ {\widehat{K}}^{-1}_{L'S'a';\ LSa}(\Ecm)
  \ q_{{\rm cm},a}^{-L-\frac{1}{2}},
\label{eq:Keffrange}
\eeq
where ${\widehat{K}}^{-1}_{L'S'a';\ LSa}(\Ecm)$ is a real, symmetric, and analytic 
function of the center-of-momentum energy $\Ecm$.
The effective range expansion given in Eq.~(\ref{eq:Keffrange}) suggests the convenience
of writing
\beq
 K^{-1}_{L'S'a';\ LSa}(\Ecm)=u_{a'}^{-L'-\frac{1}{2}}\ {\widetilde{K}}^{-1}_{L'S'a';\ LSa}(\Ecm)
  \ u_a^{-L-\frac{1}{2}},
\eeq
since ${\widetilde{K}}^{-1}_{L'S'a';\ LSa}(\Ecm)$ is real and symmetric and expected to behave
smoothly with energy $\Ecm$.  It is then straightforward to show that the quantization
condition of Eq.~(\ref{eq:quant1}) can be written
\beq
\det(1-B^{(\Pvec)}\widetilde{K})=\det(1-\widetilde{K}B^{(\Pvec)})=0,
\label{eq:quant2}
\eeq
where we define the \textit{box matrix} by
\beqy
 && \me{J'm_{J'}L'S'a'}{B^{(\Pvec)}}{Jm_JLS a} =
-i\delta_{a'a}\delta_{S'S} \ u_a^{L'+L+1}\ W_{L'm_{L'};\ Lm_L}^{(\Pvec a)}  \nn\\
&&\qquad\qquad \times\langle J'm_{J'}\vert L'm_{L'},Sm_{S}\rangle
\langle Lm_L,Sm_S\vert Jm_J\rangle.
\label{eq:Bmatdef}
\eeqy
This box matrix $B^{(\Pvec)}$ is Hermitian for $u_a^2$ real.  Whenever
$\det \widetilde{K}\neq 0$, which is usually true in the presence of interactions,
the quantization condition can also be written
\beq
  \det(\widetilde{K}^{-1}-B^{(\Pvec)})=0.
\label{eq:quant3}
\eeq
The Hermiticity of $B^{(\Pvec)}$ and the fact that $\widetilde{K}$ is
real and symmetric for real $u_a^2$ ensures that the determinants in the 
quantization conditions of Eqs.~(\ref{eq:quant2}) and (\ref{eq:quant3}) 
are real.  Note that $\widetilde{K}$ and $B^{(\Pvec)}$ do not commute in
general, which means $1-B^{(\Pvec)}\widetilde{K}$ and 
$1-\widetilde{K}B^{(\Pvec)}$ are not Hermitian.  However, it is easy
to show that their determinants must be real.

Again, rotational invariance of the $K$-matrix implies that $\widetilde{K}$ has the form
\beq 
  \langle J'm_{J'}L^\prime S^\prime a'\vert\ \widetilde{K}
\ \vert Jm_JLS  a\rangle = \delta_{J'J}\delta_{m_{J'}m_J}
 \ {\cal K}^{(J)}_{L'S'a';\ LS a}(\Ecm).
\label{eq:Ksmooth}
\eeq
When $S=S'=0$, then $J=L$ and $J'=L'$ yielding
\beq 
  \langle J'm_{J'}L^\prime 0 a'\vert\ \widetilde{K}
\ \vert Jm_JL0  a\rangle = \delta_{J'J}\delta_{m_{J'}m_J}\delta_{J'L'}\delta_{JL}
 \ {\cal K}^{(L)}_{a';\ a}(\Ecm).
\label{eq:Ksmoothb}
\eeq
Given that $\widetilde{K}^{-1}$ is expected to be analytic in $\Ecm$,
an obvious parametrization of the inverse of the $\widetilde{K}$-matrix 
over a small range of energies is using a symmetric matrix of polynomials 
in $\Ecm$:
\beq
 {\cal K}^{(J)-1}_{\alpha\beta}(\Ecm) = \sum_{k=0}^{N_{\alpha\beta}} 
c_{\alpha\beta}^{(Jk)} \Ecm^k,
\label{eq:Kfit1}
\eeq
where $\alpha,\beta$ are compound indices referring to orbital momentum $L$, total spin $S$, and
channel $a$, and the $c_{\alpha\beta}^{(Jk)}$ form a real symmetric matrix for each $k$.
Another common parame\-trization (see, for example, Ref.~\cite{Chung:1995dx})
expresses the $\widetilde{K}$-matrix as a sum of poles with a 
background described by a symmetric matrix of polynomials:
\beq
  {\cal K}^{(J)}_{\alpha\beta}(\Ecm) = \sum_{p} \frac{g_\alpha^{(Jp)} g_\beta^{(Jp)}}{\Ecm^2-m_{Jp}^2}
     + \sum_k d_{\alpha\beta}^{(Jk)} \Ecm^k,
\label{eq:Kfit2}
\eeq
where the couplings $g_\alpha^{(Jp)}$ are real and the background coefficients
$d_{\alpha\beta}^{(Jk)}$ form a real symmetric matrix for each $k$.  These can be written
in Lorentz invariant form using $\Ecm=\sqrt{s}$, where the Mandelstam variable $s=(p_1+p_2)^2$,
with $p_j$ being the four-momentum of particle $j$.

\section{Block diagonalization}

So far, we have expressed the matrices $F^{(\Pvec)}$ and $B^{(\Pvec)}$
in terms of the basis states labelled by $\ket{Jm_JLS a}$.  In this basis, the quantization 
condition in each of Eqs.~(\ref{eq:quant1}), (\ref{eq:quant2}) and (\ref{eq:quant3}) is 
difficult to use since the determinant of an infinite matrix must be evaluated.  By
transforming to a basis in which both $B^{(\Pvec)}$ and $\widetilde{K}$ 
are block diagonal, we can focus on the determinant separately in each block.
Each block has infinite dimension, but by truncating in the orbital angular momentum, 
keeping only states with $L\leq L_{\rm max}$, each truncated block has a finite and 
reasonably small size.

Under a symmetry transformation $G$ which is either an ordinary spatial rotation $R$ or 
spatial inversion $I_s$, the total momentum $\Pvec$ changes to $G\Pvec$, and if we define a 
unitary matrix $Q^{(G)}$ by
\beq
      \me{J'm_{J'}L'S'a'}{Q^{(G)}}{ Jm_JLS a}
   = \left\{\begin{array}{ll}
 \delta_{J'J}\delta_{L'L}\delta_{S'S}\delta_{a'a}
     D^{(J)}_{m_{J'}m_{J}}(R),  & (G=R),\\[4pt]
  \delta_{J'J}\delta_{m_{J'}m_J}\delta_{L'L}\delta_{S'S}\delta_{a'a}
     (-1)^{L}, & (G=I_s),
 \end{array}\right.
\label{eq:Qsymdef}
\eeq
where $D^{(J)}_{m'm}(R)$ are the familiar Wigner rotation matrices,
one can show that the box matrix satisfies
\beq
   B^{(G\Pvec)} = Q^{(G)}\ B^{(\Pvec)}\ Q^{(G)\dagger}.
\label{eq:Brotate}
\eeq
If $G$ is an element of the little group of $\Pvec$, then $G\Pvec=\Pvec$ and 
$G\svec_a=\svec_a$,  and we have
\beq
   B^{(\Pvec)} = Q^{(G)}\ B^{(\Pvec)}\ Q^{(G)\dagger},
 \qquad\mbox{($G$ in little group of $\Pvec$).}
\label{eq:Breduce2}
\eeq
Since $Q^{(G)}$ is unitary, this implies that the $B^{(\Pvec)}$ matrix commutes 
with the matrix $Q^{(G)}$ for all $G$ in the little group of $\Pvec$.
This means that we can simultaneously diagonalize $B^{(\Pvec)}$ and $Q^{(G)}$.  
By rotating into a basis formed by the eigenvectors of $Q^{(G)}$, we can reduce 
the $B^{(\Pvec)}$ matrix into a block diagonal form since the matrix elements
of $B^{(\Pvec)}$ between different eigenvectors of $Q^{(G)}$ must vanish.

Rotations, reflections, and spatial inversion do not change $J,L,S,a$ when
acting on basis state $\ket{Jm_JLSa}$.  These symmetry operations only mix
states of different $m_J$. To block diagonalize $B^{(\Pvec)}$, we apply a 
particular unitary change of basis:
\beq
  \ket{\Lambda\lambda n JLS a}= \sum_{m_J} c^{J\eta;\,\Lambda\lambda n}_{m_J} 
 \ket{Jm_JLS a},
 \label{eq:bdtrans}
\eeq
where $\eta=(-1)^L$.  The new basis vectors can be labelled by the irreducible 
representation (irrep) $\Lambda$ and irrep row $\lambda$ of the little group, 
and an integer $n$ identifying each occurrence of the irrep $\Lambda$ in the 
$\ket{Jm_JLS a}$ reducible representation.
Our procedure for computing the transformation coefficients is described
in Ref.~\cite{Morningstar:2017spu}.

Expressing the box matrix in this basis, one can show that 
$B^{(\Pvec)}$ is diagonal in $\Lambda,\lambda$, but not in the occurrence 
index $n$. Given Eq.~(\ref{eq:Bmatdef}), we find that we can write
\beq
 \me{\Lambda'\lambda' n'J'L'S' a'}{B^{(\Pvec)}}{\Lambda\lambda nJLS a}
 = \delta_{\Lambda'\Lambda}\delta_{\lambda'\lambda}\delta_{S'S}
 \delta_{a'a}\ B^{(\Pvec\Lambda_B Sa)}_{J'L'n';\ JLn}(E).
\label{eq:Bbdform}
\eeq
Notice that in Eq.~(\ref{eq:Bbdform}) we use the irrep label $\Lambda_B$ instead of 
$\Lambda$ to label the matrix elements of $B^{(\Pvec)}$.  We wish to reserve the 
irrep $\Lambda$ to describe the symmetry of the block in question for the full system, 
which includes the intrinsic parities of the constituent particles.  The $B^{(\Pvec)}$ 
matrices transform independently of these intrinsic parities.  The relationships of 
$\Lambda_B$ to $\Lambda$ when $\eta^P_{1a}\eta^P_{2a}=-1$ are summarized in 
Table~1 of Ref.~\cite{Morningstar:2017spu}.  

We have determined expressions in terms of the RGL shifted zeta functions 
for all box matrix elements with $L\leq 6$, total spin $S\leq 2$, and total 
momentum $\Pvec=(0,0,0), (0,0,p)$, as well as all box matrix elements with 
$L\leq 6$, $S\leq \frac{3}{2}$, and $\Pvec=(0,p,p), (p,p,p)$, with $p>0$.
We have developed and tested software, written in C++, to evaluate these
box matrix elements.  This software is freely available\cite{gitavail}.

Lastly, we need to express $\widetilde{K}$ in the new basis.
Given Eq.~(\ref{eq:Ksmooth}) and the orthonormality of the states
in both the $\ket{Jm_JLSa}$ basis and the block diagonal 
$\ket{\Lambda\lambda nJLSa}$ basis, one can show that
\beq
 \me{\Lambda'\lambda' n'J'L'S' a'}{\widetilde{K}}{\Lambda\lambda nJLS a}
= \delta_{\Lambda'\Lambda}\delta_{\lambda'\lambda}\delta_{n'n} \delta_{J'J}
\ {\cal K}^{(J)}_{L'S'a';\ LS a}(\Ecm),\quad (\eta=\eta'),
\label{eq:Kbdbasis}
\eeq
where $\eta=(-1)^L$ and $\eta'=(-1)^{L'}$.
If $\eta=-\eta^\prime$, the situation is much more complicated. However, in QCD,
we should never need such matrix elements.  

The box matrix is diagonal in total spin $S$ and in the compound index $a$.
However, the $\widetilde{K}$-matrix allows mixings between different spins
and channels. Thus, for a given
$\Pvec$, we can label the quantization blocks of $1-B^{(\Pvec)}\widetilde{K}$ and 
$\widetilde{K}^{-1}-B^{(\Pvec)}$ in the $\vert\Lambda\lambda n JLSa\rangle$ basis 
solely by the irrep label $\Lambda$, where $\Lambda$ is the irrep associated
with the $K$-matrix.

\section{Fitting}

Let $\kappa_j$, for $j=1,\dots,N_K$, denote the parameters that appear in the matrix 
elements of either the $\widetilde{K}$-matrix or its inverse $\widetilde{K}^{-1}$.  
Once a set of energies for a variety of two-particle interacting
states is determined, the primary goal is then to determine the best-fit estimates of the
$\kappa_j$ parameters using the quantization determinant, as well as to determine 
the uncertainties in these estimates. 

One method, which we call the \textit{spectrum method}, is described in 
Ref.~\cite{Morningstar:2017spu}, but it is very difficult to implement.
A much simpler method, known as the \textit{determinant residual method},
is advocated in Ref.~\cite{Morningstar:2017spu}. In this method, we introduce the
quantization determinant itself as a residual in the correlated $\chi^2$ to be minimized.  In the 
determinant, we use the observed box matrix elements, which requires the observed 
energies and the observed values for the particle masses, lattice size, and anisotropy.  

Expressing the quantization condition in terms of a vanishing determinant
is just a convenient way of stating that one eigenvalue becomes zero.
The determinant itself is not a good quantity to use as an observable since
it can become very large in magnitude for larger matrices.  Determinants
are susceptible to round off errors, which can make numerical minimization 
difficult. Instead of the determinant, we express the quantization condition 
using the following
function of matrix $A$, having real determinant, and scalar $\mu\neq 0$:
\beq
   \Omega(\mu,A)\equiv \frac{\det(A)}{\det[(\mu^2+AA^\dagger)^{1/2}]}.
\eeq
When one of the eigenvalues of $A$ is zero, this function is also zero. 
This function can be evaluated as a product of terms, one for each
eigenvalue of $A$. For eigenvalues of $A$ which are much smaller in magnitude than 
$\vert\mu\vert$, the associated term in the product tends towards the 
eigenvalue itself, divided by $\vert\mu\vert$.  However, the key feature 
of this function is that for eigenvalues which are much larger than 
$\vert\mu\vert$, the associated term in the product goes to $e^{i\theta}$ 
for real $\theta$.  This function replaces the large unimportant 
eigenvalues with unimodular quantities so that the function does not grow 
with increasing matrix size.  This is a much better behaved function, 
bounded between -1 and 1 when the determinant is real, which still reproduces 
the quantization condition.  The constant $\mu$ can be chosen to optimize ease 
of numerical root finding or $\chi^2$ minimization.  In this study, we chose
$\mu$ by starting with $\mu=1$, then increasing $\mu$ until the $\chi^2$
value at the minimum did not change very much as $\mu$ was further increased.

In this method, the model fit parameters are just the $\kappa_i$ parameters,
and the residuals are chosen to be
\beq
    r_k = 
          \Omega\Bigl(\mu, 1-  B^{(\Pvec)}(E_{{\rm cm},k}^{({\rm obs})})
    \ \widetilde{K}(E_{{\rm cm},k}^{({\rm obs})})\Bigr), 
 \qquad  (k=1,\dots,N_E),
\eeq
or the matrix $\widetilde{K}(E_{{\rm cm},k}^{({\rm obs})})^{-1}-
B^{(\Pvec)}(E_{{\rm cm},k}^{({\rm obs})})$ could be used in the $\Omega$ function.
Clearly, the model predictions in this method are dependent on the observations 
themselves, so the covariance of the residual estimates must be recomputed and 
inverted by Cholesky decomposition throughout the minimization as the $\kappa_j$ 
parameters are adjusted.  However, this is still much simpler than the root 
finding required in the spectrum method.  

An advantage of this method is that 
the complicated RGL zeta functions only need to be computed for the box matrix 
elements as observables; they do not need to be recomputed as model
parameters are changed.

\section{Tests of fitting procedures}

\begin{table}[t]
\caption[Fitting]{First tests of the determinant residual method applied to the
interacting $\pi\pi$ energies in the $I=1$ nonstrange channel described in
Ref.~\cite{Bulava:2016mks}.  These energies were obtained on a $32^3\times 256$
anisotropic lattice with $m_\pi\approx 240$~MeV.  In Ref.~\cite{Bulava:2016mks},
the number of energy levels used was $N_E=19$.  The fits below use $N_E=20$
by including an additional energy from a $B_{1}^+\ d^2=1$ irrep in
which the leading partial wave is $L=3$. The fits shown
used $\Omega(\mu,\widetilde{K}^{-1}-B)$ as the residuals.
\label{tab:tests}}
\begin{center}
\begin{tabular}{ccccccl}
\hline
$\mu$ &  $N_E$ & $m_{\rho}/m_{\pi}$  & $g$ &  $m_{\pi}^{7}a_3$    
  & $m_{\pi}^{11} a_5 $ & $\chi^2/\mathrm{dof}$ \\ 
\hline
1   &  20 &  3.338(13)   & 5.91(17)  & 0.0001(12) & -0.00016(11) & 1.75 \\ 
2   &  20 &  3.341(20)   & 5.91(22)  & 0.0001(15) & -0.00020(14) & 1.43 \\ 
4   &  20 &  3.345(24)   & 5.92(26)  & 0.0001(17) & -0.00018(19) & 1.26 \\ 
8   &  20 &  3.348(26)   & 5.96(28)  & -0.0001(12) & -0.00010(24) & 1.18 \\ 
10   & 20 &   3.349(26)   & 5.97(27)  & -0.0002(11) & -0.00007(25) & 1.16 \\ 
12   &  20 &  3.349(25)   & 5.97(27)  & -0.00021(100) & -0.00006(24) & 1.15 \\ 
\hline
\end{tabular}
\end{center}
\end{table}

As first tests, we applied the determinant residual method to the interacting $\pi\pi$ 
energies in the $I=1$ nonstrange sector in irreps relevant for extracting the
$P$-wave amplitude.  The operators used and the energies obtained are 
described in Ref.~\cite{Bulava:2016mks}. These energies were obtained on a $32^3\times 256$
anisotropic lattice with $m_\pi\approx 240$~MeV.  
Defining
$
   k_0=2\pi/(m_\pi L),
$
the fit forms we used are
\beq
(\widetilde{K}^{-1})_{11} = \frac{6\pi\Ecm}{k_0^3m_{\pi} g^2}
\left(\frac{m_{\rho}^2}{m_{\pi}^2} - \frac{\Ecm^2}{m_{\pi}^2} \right), \quad
(\widetilde{K}^{-1})_{33} = \frac{1}{k_0^7 m_{\pi}^7 a_3},\quad
(\widetilde{K}^{-1})_{55} = \frac{1}{k_0^{11} m_{\pi}^{11} a_5}.
\eeq

Some of our results are presented in Table~\ref{tab:tests}. In Ref.~\cite{Bulava:2016mks},
the number of energy levels used was $N_E=19$.  In the fits shown in Table~\ref{tab:tests}, 
we also included an elastic energy from an additional $B_{1}^+\ d^2=1$ irrep in which the 
leading partial wave is $L=3$.   In the fits listed, we
utilized $\Omega(\mu,\widetilde{K}^{-1}-B)$ as the residuals.  Using the
$\Omega$ function, we were able to find the minimum of the $\chi^2$ function much
more easily.  For $\mu=1$, we found that the minimum $\chi^2/{\rm dof}$ values were
uncomfortably large.  This was remedied by increasing $\mu$ to a value
around $\mu=8$ or larger.

The most important thing that the test fits in Table~\ref{tab:tests} demonstrate 
is that the effects of higher partial waves can be taken into account using the 
determinant residual method.  Also, our results show that the phase shifts from the 
$L=3$ and $L=5$ waves are negligible in this energy range, justifying our neglect 
of these waves in Ref.~\cite{Bulava:2016mks}.  This is consistent with a 
phenomenological determination of $m_\pi^7a_3=5.65(21)\times 10^{-5}$ taken from 
Ref.~\cite{GarciaMartin:2011cn}.  

In the future, we plan to utilize both the spectrum and residual determinant
methods in the analysis of meson-meson and meson-baryon systems
involving multiple channels.  Studies involving the $K^\ast(892)$ and
$a_0(980)$ should appear soon.  Various baryon resonances are also being
investigated.

\section{Conclusion}

This talk reported on work completed in Ref.~\cite{Morningstar:2017spu} to 
provide explicit formulas, software, and new fitting implementations 
for carrying out two-particle scattering studies using energies
obtained from lattice QCD. 
We introduced a so-called ``box matrix'' $B$ which describes how the partial
waves fit into the cubic finite volumes of lattice QCD simulations.  The
quantization condition was expressed in terms of this Hermitian matrix $B$ and
the real, symmetric scattering $K$-matrix.  The effective range expansion was
used to introduce a smooth, well-behaved matrix $\widetilde{K}^{-1}$.  We obtained explicit
expressions for the box matrix elements for several spins and center-of-momentum 
orbital angular momenta up to $L=6$ with total momentum of the form 
$\Pvec=(0,0,0), (0,0,p), (0,p,p), (p,p,p)$ for $p>0$.  More importantly,
the software for evaluating all of these box matrix elements was made available.
Lastly, we described a fitting strategy for estimating the
parameters used to approximate the $\widetilde{K}$-matrix.    First tests 
involving $\rho$-meson decay to two pions included the $L=3$ and $L=5$ partial 
waves, and the contributions from these higher waves were found to be negligible 
in the elastic energy range.

\vspace{4mm}
\noindent
{\bf Acknowledgements}: 
This work was supported by the U.S.~National Science Foundation 
under award PHY-1613449.  Computing resources were provided by
the Extreme Science and Engineering Discovery Environment (XSEDE)
under grant number TG-MCA07S017.  XSEDE is supported by National 
Science Foundation grant number ACI-1548562. We acknowledge helpful 
conversations with Raul Briceno and Steve Sharpe.

\bibliography{cited_refs}

\end{document}